\documentstyle[prb,preprint,aps,psfig]{revtex}    

\begin{document}

\draft
\def\fiwi{15cm}
\newcommand{\no}{\nonumber}
\newcommand{\beq}{\begin{equation}}
\newcommand{\eeq}{\end{equation}}
\newcommand{\beqa}{\begin{eqnarray}}
\newcommand{\eeqa}{\end{eqnarray}}

\title{Self-consistent theory of transport in superconducting wires}

\author{J. S\'anchez-Ca\~nizares and F. Sols}
\bigskip
\address{
Departamento de F\'{\i}sica Te\'orica de la Materia Condensada, C-V,
and\\
Instituto Universitario de Ciencia de Materiales ``Nicol\'as
Cabrera''\\
Universidad Aut\'onoma de Madrid, E-28049 Madrid, Spain}

\maketitle
\begin{abstract}

We study superconducting transport in homogeneous wires in the
cases of both equilibrium and nonequilibrium quasiparticle
populations, using the quasiclassical Green's function technique.
We consider superconductors with arbitrary current densities and
impurity concentrations ranging from the clean to the dirty limit.
Local current conservation is guaranteed by ensuring that the
order parameter satisfies the self-consistency equation at each
point. For equilibrium transport, we compute the current, the
order parameter amplitude, and the quasiparticle density of states
as a function of the superfluid velocity, temperature, and
disorder strength. Nonequilibrium is characterized by incoming
quasiparticles with different chemical potentials at each end of
the superconductor. We calculate the profiles of the
electrostratic potential, order parameter, and effective
quasiparticle gap. We find that a transport regime of
current-induced gapless superconductivity can be achieved in clean
superconductors, the stability of this state being enhanced by
nonequilibrium.


\end{abstract}

\vspace{.5cm}



\def\dvsvs{
\begin{figure}
\psfig{file=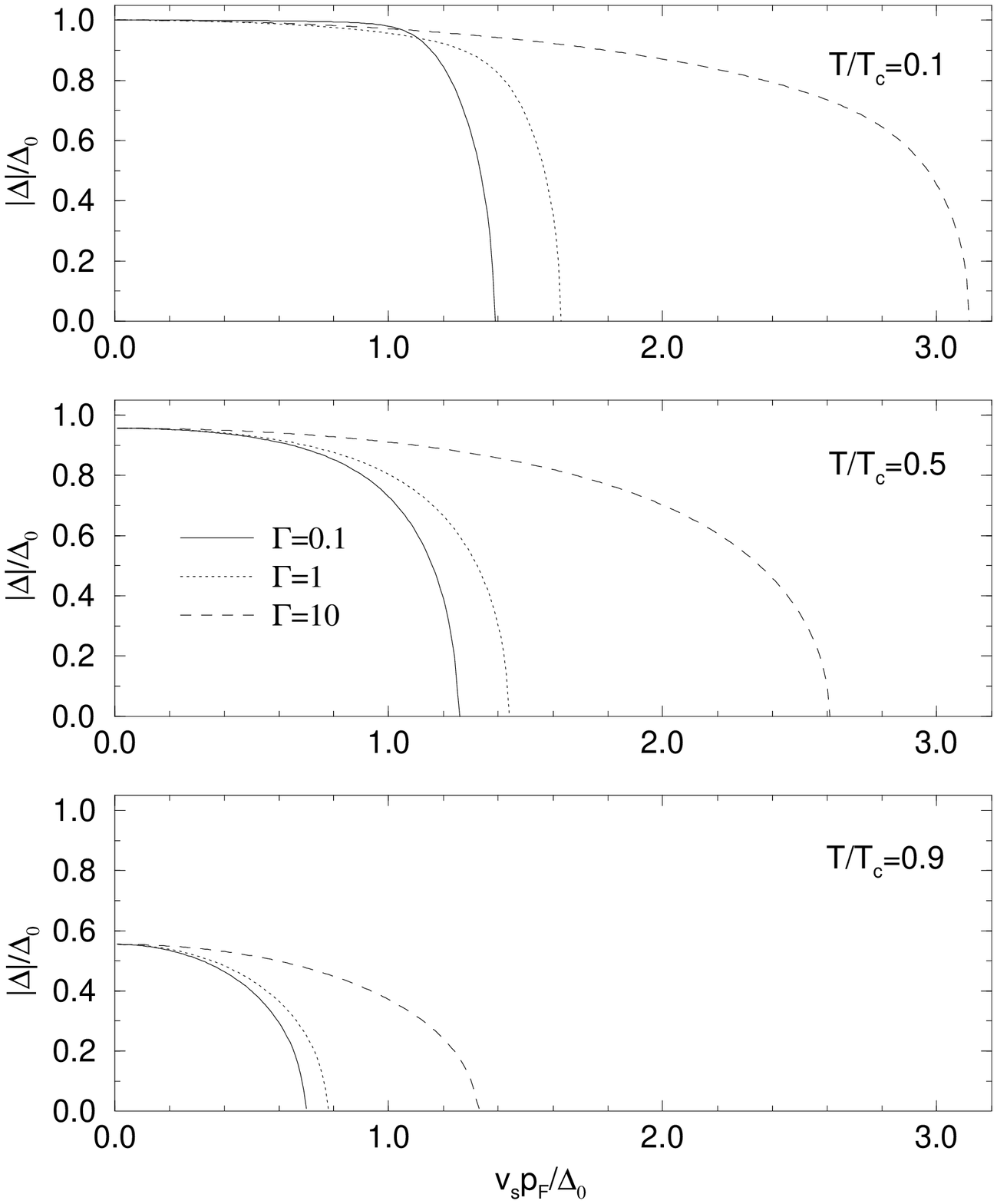,width=\fiwi,angle=0} \caption{\label{dvsvs}
Impurity averaged order parameter amplitude as a function of the
superfluid velocity, for different values of the normalized
disorder rate $\Gamma=\hbar /\tau\Delta_0$ and temperature. }
\end{figure}
}

\def\jcvsvs{
\begin{figure}
\psfig{file=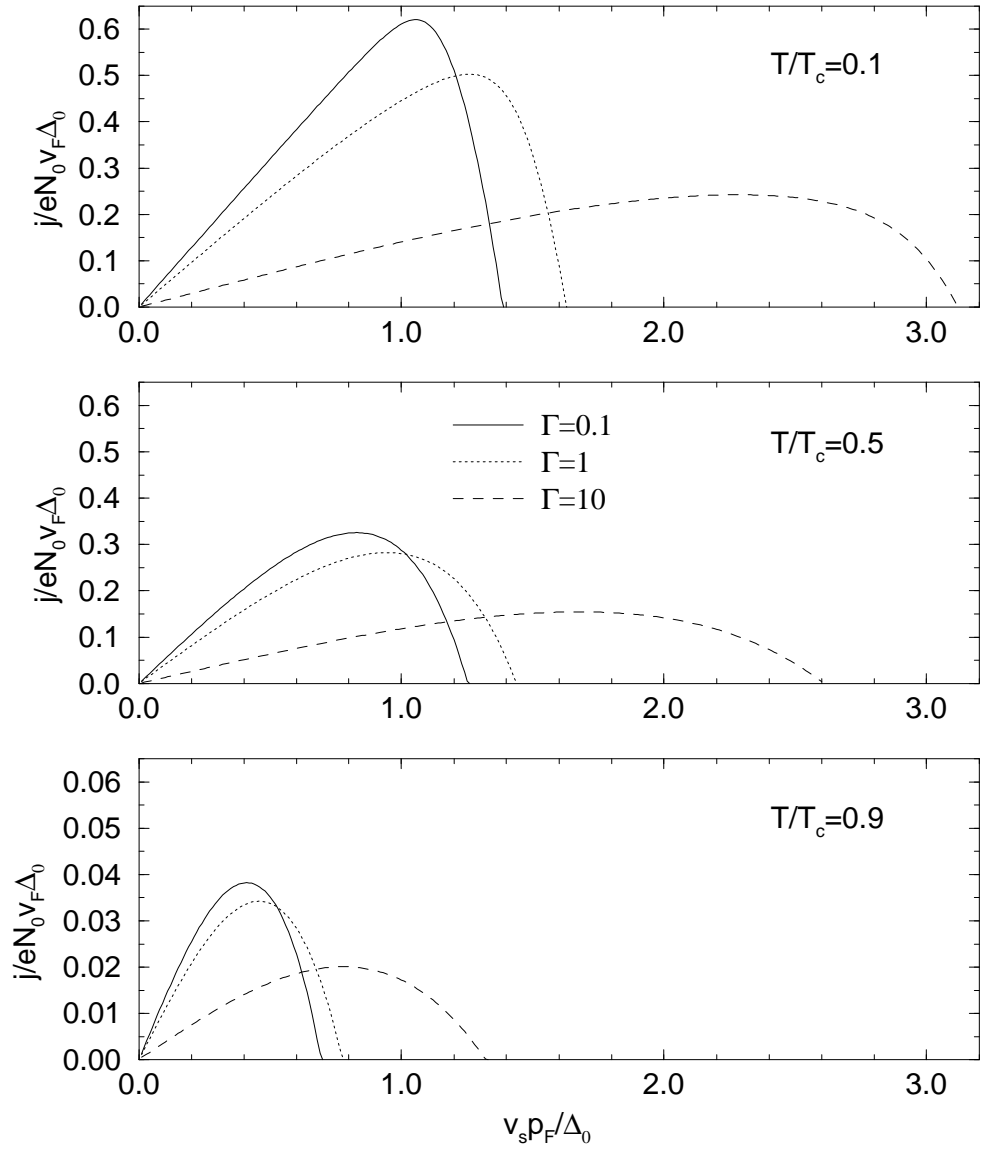,width=\fiwi,angle=0} \caption{\label{jcvsvs}
Same as Fig. \ref{dvsvs} for the current density. }
\end{figure}
}

\def\cmvsdt{
\begin{figure}
\psfig{file=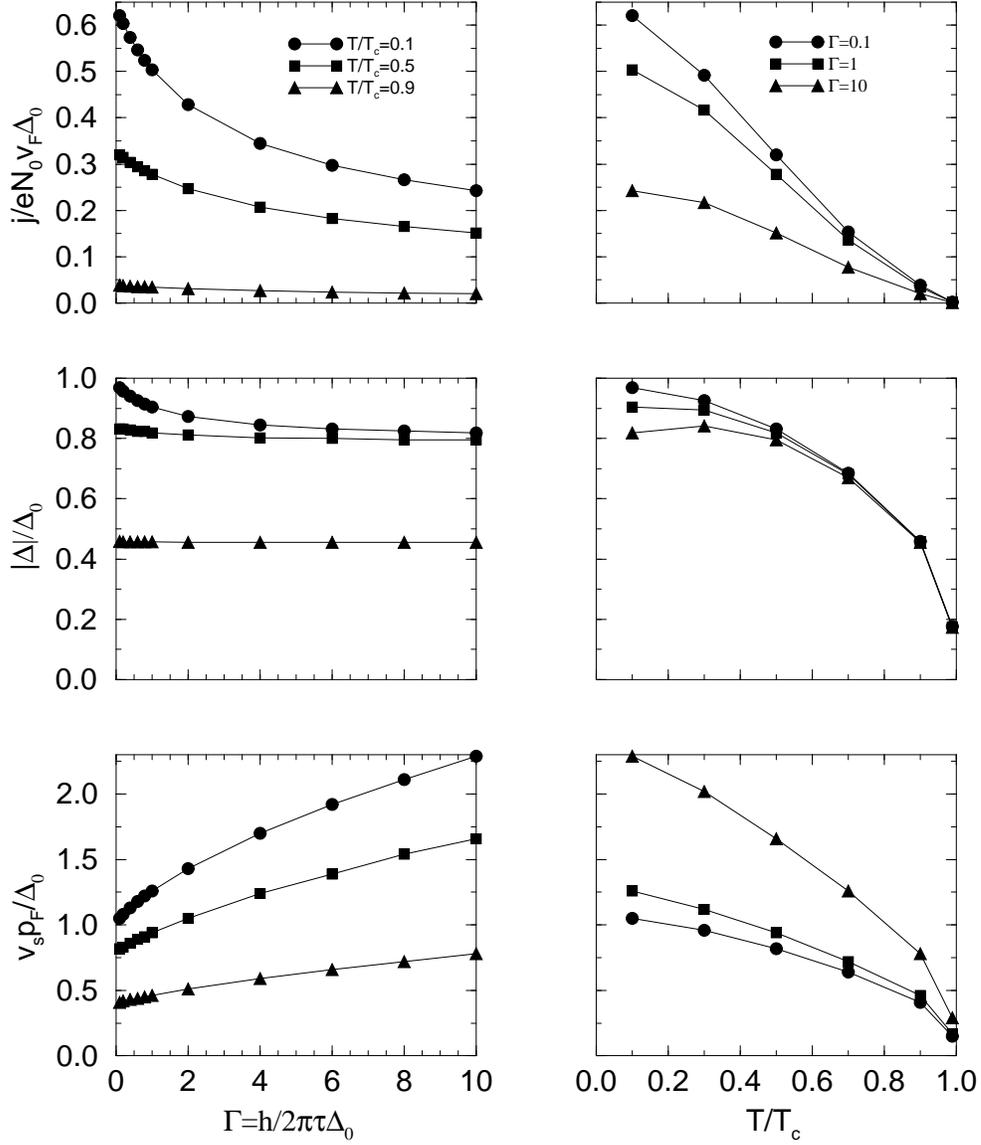,width=\fiwi,angle=0}
\caption{\label{cmvsdt} Impurity averaged {\it critical} magnitudes as a function
of the disorder (left) and of the the temperature (right).
}
\end{figure}
}

\def\jcvsd{
\begin{figure}
\psfig{file=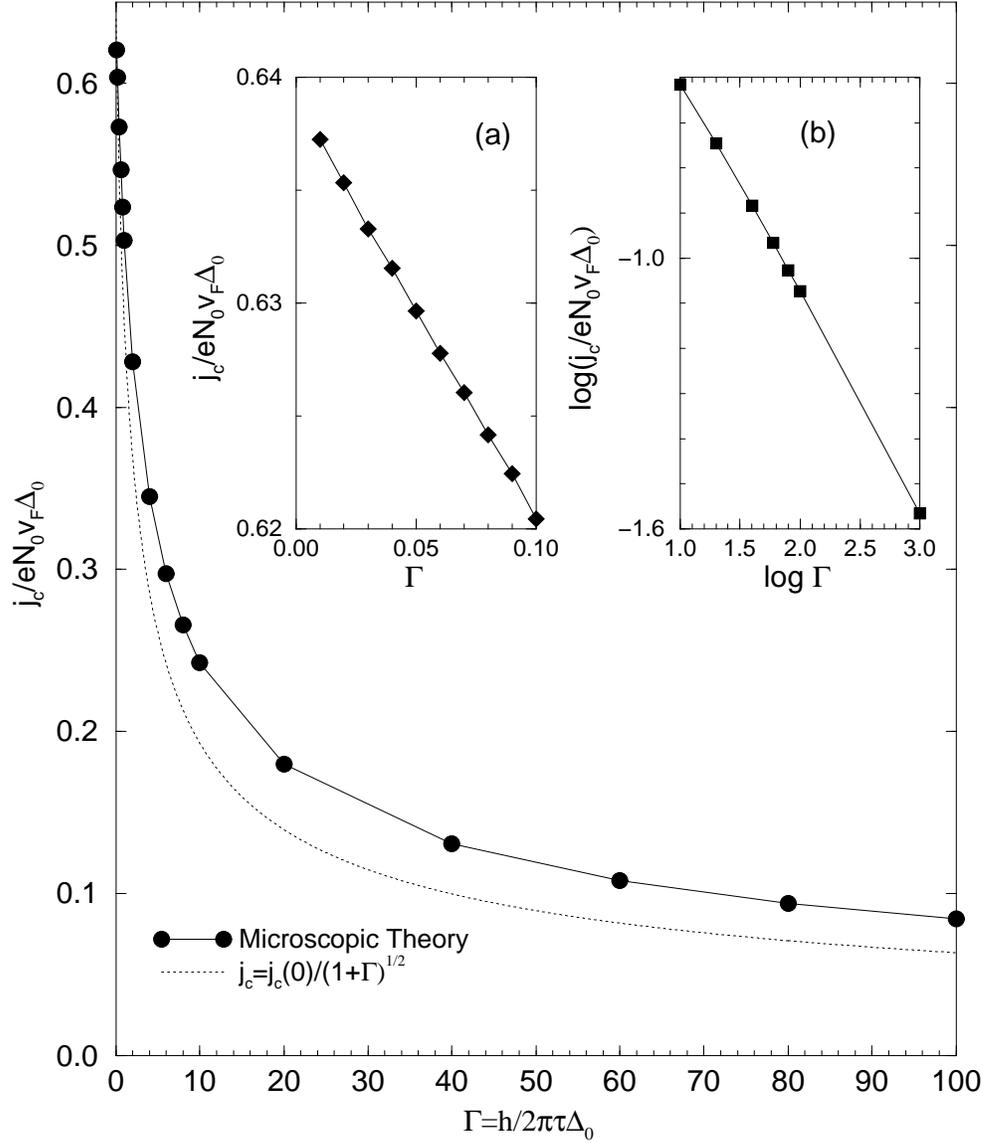,width=\fiwi,angle=0} \caption{\label{jcvsd}
Impurity averaged critical current density as a function of the
disorder for $T=0.1T_c$. Insets: (a) Ballistic case ($\Gamma \ll
1$): Linear decay of the critical current with the disorder rate.
(b) Diffusive case ($\Gamma \gg 1$): Power-law $j_c \propto \Gamma
^{-\beta}$. We find $\beta \approx 0.47$, slightly different from
the simple prediction of $0.5$ (dotted line). }
\end{figure}
}

\def\dos{
\begin{figure}
\psfig{file=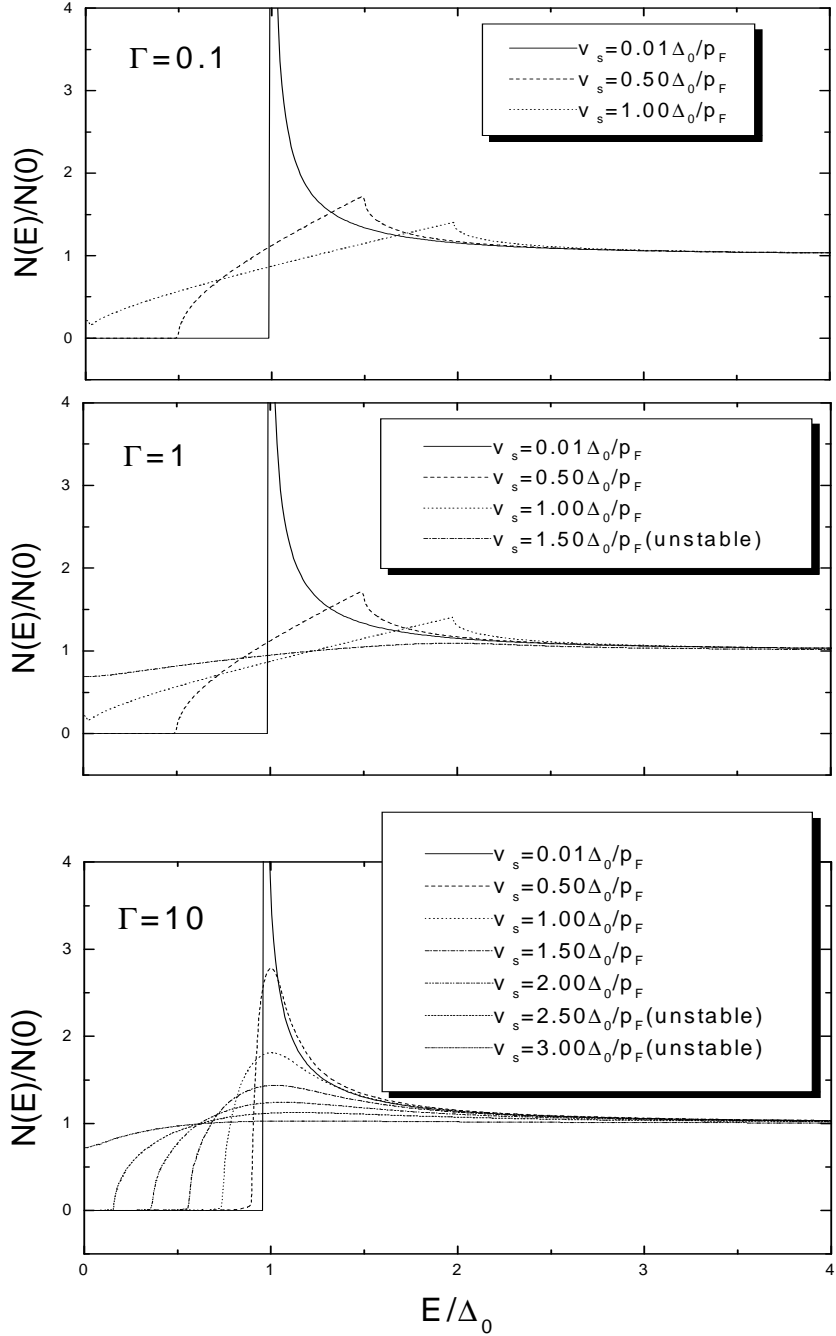,width=\fiwi,angle=0} \caption{\label{dos}
Impurity averaged density of states for several superfluid
velocities and disorder strengths. All results are given for
$T=0.1T_c$. }
\end{figure}
}

\def\nsnpic{
\begin{figure}
\psfig{file=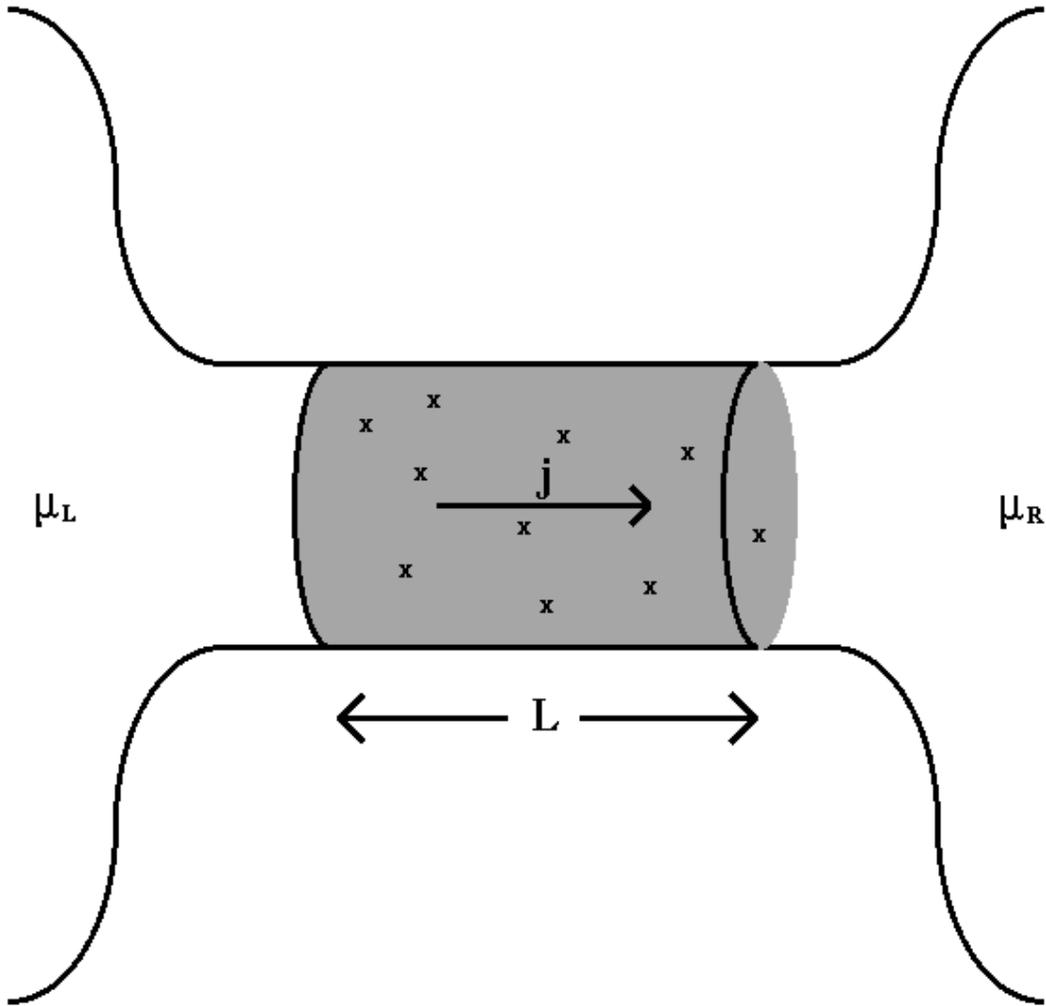,width=\fiwi,angle=0} \caption{\label{nsnpic}
Schematic representation of a typical NSN structure, where S is a
dirty superconductor. The normal leads are assumed to be perfect
and connected to wide reservoirs through ideal contacts. Grey
colour marks the superconducting zone of length $L$. Small crosses
signal possible impurity positions. The picture shows a specific
realization of disorder within the superconductor. }
\end {figure}
}

\def\jdvsv{
\begin{figure}
\psfig{file=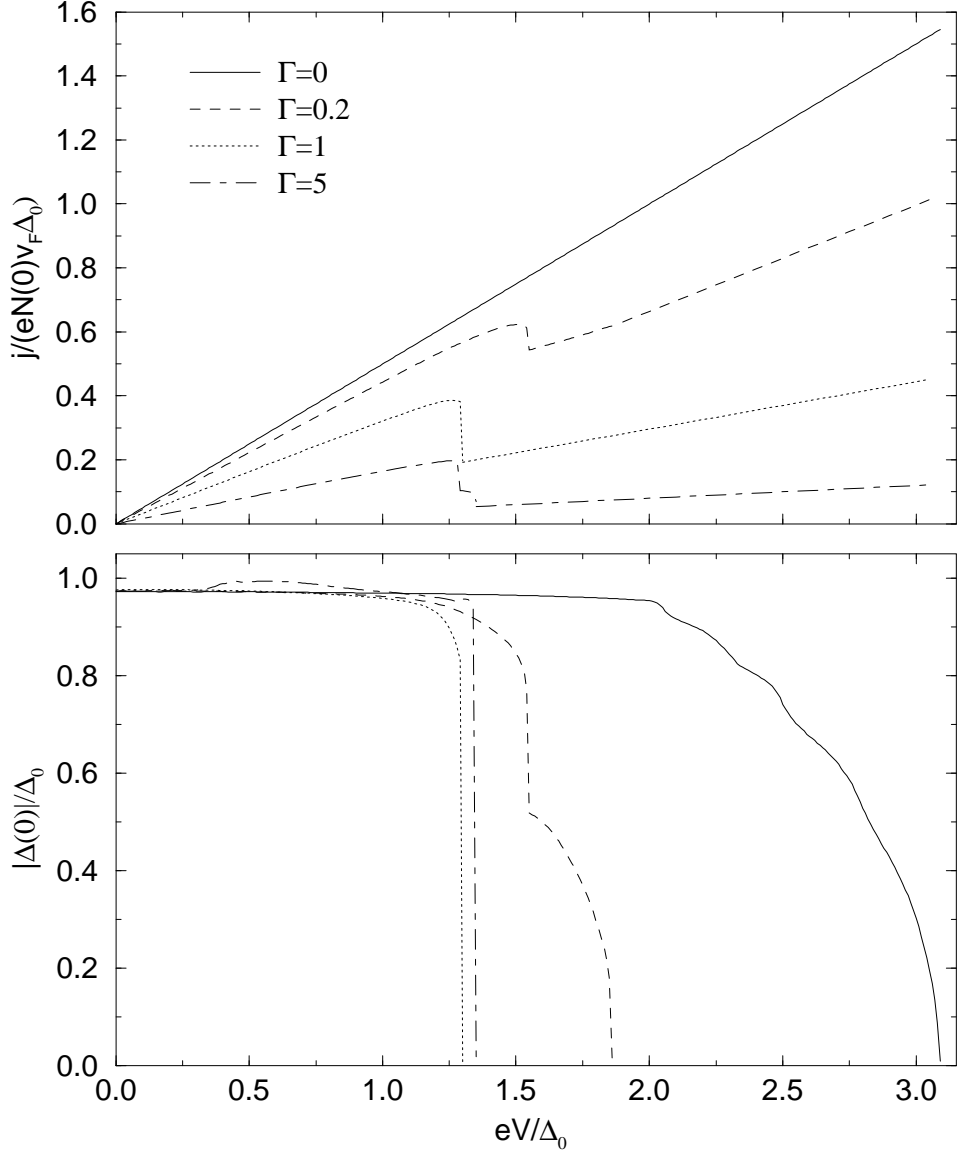,width=\fiwi,angle=0} \caption{\label{jdvsv}
Top: Current density as a function of the applied voltage in the
structure of Fig. \ref{nsnpic}, for different values of the
disorder. Bottom: Order parameter amplitude at the center of the
superconductor, for the same values of $\Gamma$. Here, $L=3\xi_0$
and $T=0.01T_c$. }
\end {figure}
}

\def\vddnvsz{
\begin{figure}
\psfig{file=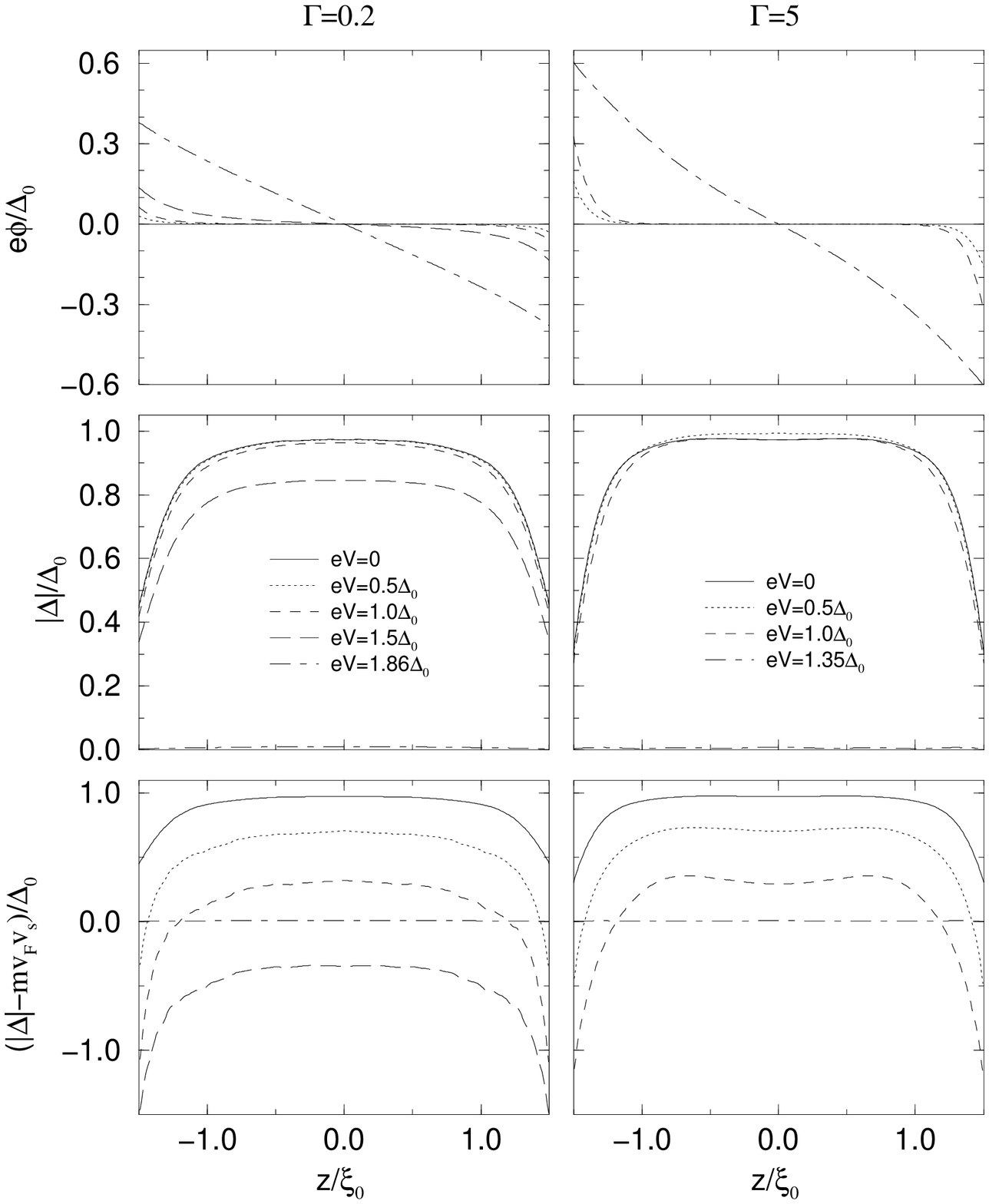,width=\fiwi,angle=0}
\caption{\label{vddnvsz} Comparison between ballistic (left) and
diffusive (right) limits in the structure of Fig. \ref{nsnpic}.
Top pannels: Profile of the electrostatic potential within the
superconductor. Center pannels: Order parameter amplitude. Bottom
pannels: Threshold $\Delta_-\equiv|\Delta|-\hbar v_F\varphi
'/2=|\Delta|-mv_Fv_s$. At positions where $\Delta_-\!<\!0$ the
superconductor is gapless. \cite{comm4} $L$ and $T$ as in Fig.
\ref{jdvsv}}
\end {figure}
}

\section{Introduction}

The quasiclassical theory of superconductivity was proposed by
Eilenberger \cite{eile68} and Larkin and Ovchinnikov,
\cite{lark69} and later developed by Usadel \cite{usad70} and
Eliashberg. \cite{elia71} It is especially useful to study
transport in systems with characteristic length scales much
greater than the Fermi wave length $\lambda_F$, which is
effectively integrated out of the problem. In the seventies and
eighties, the theory of nonequilibrium superconductivity was
developed,\cite{gray81} and the response of the quasiparticle
distribution function to external perturbations was investigated.
Theoretical developments in Refs. \cite{schm75,arte78} were based
on the quassiclassical Green's function (QCGF) technique. The
calculation became possible of characteristic relaxation times due
to phonons and paramagnetic impurities when an excess of
quasiparticles is injected into the superconductor. In most of the
cases, temperatures near the critical temperature and disorder
strengths in the dirty limit were assumed, although the theory is
potentially valid for all ranges of disorder and temperature. The
effect of a moving condensate with a finite superfluid velocity
$v_s$ was studied in Refs.
\cite{arte78,peth79b,shel80,scho81,beye86}. In particular, the
quasiparticle relaxation time for energies between $\Delta_-$ and
$\Delta_+$, where $\Delta_{\pm}\equiv|\Delta| \pm \hbar v_s p_F$,
was calculated \cite{scho81,beye86} ($\Delta_{\pm}$ are the
direction-dependent effective quasiparticle energy gaps; $\Delta$
is the conventional gap function playing the role of
superconducting order parameter). However, the influence of $v_s$
on $|\Delta|$ through the self-consistency condition \cite{dege66}
was not analysed.

During the nineties, a strong interest has developed on the
physics of elastic, low current transport. Part of this work
\cite{lamb91} has been based on the resolution of the microscopic
Bogoliubov -- de Gennes (BdG) equations, \cite{dege66,tink96}
whose equivalence with QCGF was proved in the clean limit by Beyer
{\it et al.} \cite{beye86} The QCGF technique has allowed
theorists to explain most of the experiments involving
normal-superconductor (NS) interfaces. By permitting the study of
spectral and spatial properties with inclusion of impurity
averaged disorder, QCGF have become a most adequate tool to
understand phenomena such as zero bias anomalies \cite{kast91} and
re-entrance of the conductance, \cite{char96} including its
non-monotonic behavior as a function of temperature, voltage bias,
or phase difference between external superconductors.
\cite{lamb97} These effects are now understood as an interplay
between proximity effect and disorder. \cite{volk92,golu97}

The study of equilibrium superconducting transport for arbitrary
currents was already initiated in the sixties.
\cite{roge60,bard62,maki69} Rogers \cite{roge60} and Bardeen
\cite{bard62} developed a thermodynamic theory, and Maki
\cite{maki69} gave a unified description of the clean and dirty
limits using a Green's function method. More recently, Bagwell
\cite{bagw94} has performed a similar study for clean
superconductors within the BdG framework. The close connection
between the implementation of self-consistency in the order
parameter equation \cite{dege66} and current conservation has been
noted by several authors. \cite{bagw94,furu91,sols94} The
implication is that a self-consistent description is essential in
transport scenarios involving large supercurrents or
superconductor lengths. \cite{sols94}

The combined effect of a moving condensate and a nonequilibrium
distribution of quasiparticles has been addressed recently within
a self-consistent scheme.
\cite{sanc95,mart95,sanc96,sanc97,sanc98,bagw99} These authors
have solved the BdG equations for one-dimensional models of
ballistic transport and have predicted physical features such as
the existence of an Andreev-transmission dominated transport
regime, \cite{sanc95,sanc96,sanc97} the enhancement of excess
current due to a finite $v_s$, \cite{sanc97,bagw99} and the
possibility of current-induced gapless superconductivity (GS).
\cite{sanc95,mart95,sanc97,sanc98}

The primary purpose of this paper is to study the robustness of
these effects against the inclusion of realistic physical factors
such as scattering by impurities and the presence of many
transverse channels. Specifically, we employ the QCGF technique to
study both equilibrium and nonequilibrium transport in
superconducting wires for arbitrary currents and applied voltages.
Our equilibrium transport study complements work done by Maki
\cite{maki69} in that we compute transport properties for
different disorder strengths, ranging from the clean to the dirty
limit. In particular, we calculate the critical current density
for arbitrary disorder, thus going beyond the Usadel equations,
\cite{usad70} valid only in the dirty limit. We also assume that
the length of the superconductor is much smaller than the
inelastic scattering length.

In this work we have calculated stationary mean field solutions
describing the response of the superconductor to an externally
applied voltage. The obtained transport configurations are stable
at sufficiently low temperatures and wire widths $W$ not much
smaller than the Meissner penetration length $\lambda$. In this
regime (specifically for $W\agt \lambda/25$, according to the
estimate of Ref. \cite{zaik96}), both thermal \cite{lang67} and
quantum \cite{zaik96} phase-slips are energetically unfavored. On
the other hand, we assume quasi--one-dimensional superconducting
wires, for which the Meissner effect and transverse variations of
the order parameter can be neglected. This gives us a window of
parameters for which our theory is quantitatively valid. Despite
these considerations, we wish to point out that a satisfactory
understanding is still lacking of the crossover from stationary
configurations to dynamic phase-slips in the response to
externally applied voltages. A unified description of both
phenomena should be the object of future theoretical study.

In section II, we give a brief, self-contained presentation of the
QCGF technique (see Ref. \cite{lamb97} for an updated review on
this topic). Section III is devoted to equilibrium transport. The
dependence of the order parameter, current density, and
quasiparticle density of states on the superfluid velocity are
computed exactly, and several critical magnitudes are calculated
for different temperatures and disorder strengths. In section IV,
we study nonequilibrium transport with elastic impurity scattering
of arbitrary strength, and discuss the robustness of the GS
regime. We solve the QCGF equation of motion with boundary
conditions describing the injection of quasiparticles from normal
reservoirs. The conclusions are presented in section V.

\section{The quasiclassical Green's function}

We follow Refs. \cite{lark69,lamb97} in the derivation of the
equation for QCGF. The Dyson equation for the matrix Green's
function $\check{G}$ in the non equilibrium Keldysh formalism
\cite{keld65} is
\begin{equation}
[\check{G}_0^{-1}-\check{\Sigma}]\check{G}=\check{1}
\label{ldyson},
\end{equation}
where \beqa \check{G}\equiv \left[ \begin{array}{cc} \hat{G}^R &
\hat{G} \\ 0 & \hat{G}^A \end{array} \right], \label{gkeldysh}
\eeqa $\check{G}_0^{-1}$ being the usual, one-body, inverse matrix
Green's function, and $\check{\Sigma}$ the self-energy matrix in
Keldysh space. Following standard notation, the symbol $\hat{}\;$
indicates $2 \times 2$ matrices in Nambu space. The retarded,
advanced, and Keldysh Green's functions are defined, respectively,
as: \beq \hat{G}^R(1,2)\equiv
\theta(t_1-t_2)[\hat{G}^{<}(1,2)-\hat{G}^{>}(1,2)], \label{gr}
\eeq \beq \hat{G}^A(1,2)\equiv
-\theta(t_2-t_1)[\hat{G}^{<}(1,2)-\hat{G}^{>}(1,2)] \label{ga},
\eeq \beq \hat{G}(1,2)\equiv [\hat{G}^{<}(1,2)+\hat{G}^{>}(1,2)],
\label{gk} \eeq where \beqa i\hat{G}^{<}(1,2)\equiv \left[
\begin{array}{cc}
\!<\!\psi_{\uparrow}(1)\psi_{\uparrow}^{\dag}(2)\!>\! &
\!<\!\psi_{\uparrow}(1)\psi_{\downarrow}(2)\!>\!\\
-\!<\!\psi_{\downarrow}^{\dag}(1)\psi_{\uparrow}^{\dag}(2)\!>\!&
-\!<\!\psi_{\downarrow}^{\dag}(1)\psi_{\downarrow}(2)\!>\!
\end{array} \right],
\label{gmp}
\eeqa
\beqa
i\hat{G}^{>}(1,2)\equiv
-\left[ \begin{array}{cc}
\!<\!\psi_{\uparrow}(2)^{\dag}\psi_{\uparrow}(1)\!>\!&
\!<\!\psi_{\downarrow}(2)\psi_{\uparrow}(1)\!>\!\\
-\!<\!\psi_{\uparrow}^{\dag}(2)\psi_{\downarrow}^{\dag}(1)\!>\!&
-\!<\!\psi_{\downarrow}(2)\psi_{\downarrow}^{\dag}(1)\!>\!
\end{array} \right].
\label{gpm} \eeqa In these expressions, we have used the standard
abbreviations $1\equiv({\bf r}_1,t_1)$ and $2\equiv({\bf
r}_2,t_2)$, which allow us to rewrite Eq. (\ref{ldyson}) as \beq
\int d2[\check{G}_0^{-1}(1,2)-\check{\Sigma}(1,2)]\check{G}(2,3)=
\check{\delta}(1-3). \label{ildyson} \eeq Substracting from
(\ref{ldyson}) its conjugate equation, we obtain \beq
[\check{G}_0^{-1}-\check{\Sigma},\check{G}]=0. \label{dyson} \eeq
This equation may be simplified by going to the center-of-mass and
relative coordinates, defined as ${\bf r}_{1,2}\equiv {\bf R}\pm
{\bf r}/2$, $t_{1,2}\equiv T\pm t/2$. One Fourier transforms now
with respect to the relative variables ${\bf r}$ and $t$, and
introduces the QCGF defined by \beq \check{g}({\bf
R},T,\hat{p},E)\equiv \frac{i}{\pi} \int_{-\infty}^{\infty}d\xi
\check{G}({\bf R},T,{\bf p},E), \label{qcgf} \eeq where $\xi
\equiv p^2/2m-\mu$ is the free-electron energy measured from the
Fermi level and $\hat{p}\equiv{\bf p}/|{\bf p}|$. The key
assumption in the quasiclassical approximation is that the
self-energy $\check{\Sigma}$ is almost $\xi$-independent. Then we
can set $\xi=0$ in $\check{\Sigma}$ and derive for the stationary
case (neglecting the effect of magnetic fields), \beq \hbar
v_F\hat{p}\cdot \nabla_{{\bf R}}\check{g}=iE[\check{\tau}_3,
\check{g}]-i[\check{\Sigma},\check{g}]. \label{motioneq} \eeq
$\check{\tau}_3$ is a block-diagonal matrix with block entries
like in the third Pauli matrix $\hat{\tau_3}$. After the
subtraction procedure, a normalization condition is needed for
$\check{g}$. In Ref. \cite{shel85}, Shelankov studies the general,
nonstationary case, and \beq \check{g}\check{g}=1 \label{nc} \eeq
is shown to be a useful choice.

Together with the normalization condition (\ref{nc}), Eq.
(\ref{motioneq}) determines the QCGF. The remaining physical
quantities can be expressed in terms of $\check{g}$. Specifically,
the current density flowing through the system is
\cite{lamb97,ramm86} \beq {\bf j}=-\frac{1}{4}eN(0)v_F\int dE \int
\frac{d\hat{p}}{4\pi}\hat{p}\, \mbox{Tr}(\hat{\tau}_3 \hat{g}),
\label{curreq} \eeq $N(0)$ being the single-particle density of
states per spin at the Fermi level.

Within the BCS approximation, the pairing effect is introduced in
$\check{\Sigma}$ via the order parameter $\Delta$. The resulting
BCS self-energy has retarded and advanced components \cite{ramm86}
\beq \hat{\Sigma}_{\rm{BCS}}^{R,A}=-i[\mbox{Re}(\Delta)
\hat{\tau}_1-\mbox{Im}(\Delta) \hat{\tau}_2]. \label{sigmabcs}
\eeq In this language, the self-consistency equation for the order
parameter reads \cite{ramm86,brud90} \beq \Delta =
-\frac{i}{8}g\int \frac{d\hat{p}}{4\pi}\int_{-E_D}^ {E_D}dE \,
\mbox{Tr} [(\hat{\tau}_1-i\hat{\tau}_2)\hat{g}], \label{deltaeq}
\eeq with $g$ the electron-phonon coupling constant and $E_D$ the
usual BCS cutoff energy. Eqs. (\ref{motioneq}), (\ref{nc}),
(\ref{curreq}), and (\ref{deltaeq}) form the basic blocks for
calculations in the following sections.

\section{The homogeneous wire}

We begin by studying transport in superconducting wires with
arbitrary disorder and homogeneous on a scale much greater than
the mean free path $l$. It is well known that, up to first order
in perturbation theory, \cite{kim97} impurities are unable to
lower $T_c$ in conventional $s$-wave superconductors, the reason
being that non-magnetic impurities do not break time-reversal
symmetry. \cite{ande59,bayi98} However, the effect of impurities
is expected to be important when a non negligible superfluid
velocity is present in the system. For instance, it is known that
the critical current decreases with impurity concentration.
\cite{dege66} Moreover, the presence of a finite superflow
introduces an intrinsic anisotropy in the system which one should
expect to be sensitive to the presence of random scatterers. On
the other hand, the study of transport in even nominally clean
quasi--one-dimensional wires ($l\!\gg\!\xi_0\!\gg\! W$, $\xi_0$
being the zero temperature coherence length \cite{tink96}) must
include the effect of random disorder if scattering at the surface
is diffusive. \cite{maki69} As mentioned above, the use of the
QCGF technique permits an easy implementation of averages over
impurity configurations. This allows for quantitative comparison
with experiments and represents a notable advantage over
techniques based on the resolution of the BdG equations. There
have already been calculations exploring the extreme cases of
ballistic \cite{roge60} and diffusive (dirty) superconductors
\cite{maki69} in equilibrium. We wish to study the {\it crossover}
between these two limits. In particular, we want to calculate how
the critical current density decays with disorder and compare our
results with the predictions based on macroscopic descriptions of
the diffusive limit.

\subsection{The model and its solution}

We introduce disorder within the simplest approximation of
incoherent multiple scattering by impurities. \cite{rick80} We
average over impurity configurations compatible with a given
degree of macroscopic disorder, retaining only the leading term in
an expansion in powers of $(k_Fl)^{-1}$. \cite{maha90} Under these
rather standard assumptions, the contribution of disorder to the
self-energy may be written as \cite{lamb97,ramm86} \beq
\check{\Sigma}_{\rm{imp}}=-\frac{i\hbar}{2\tau}\!<\!\check{g}\!>\!,
\label{sigmaimp} \eeq where the brackets are meant to indicate
angular average over the Fermi surface
[$<\!f\!>\equiv\int(d\hat{p}/4\pi)f$]. The {\it scattering time}
$\tau$ coincides with the {\it transport time} $\tau_{\rm{tr}}$
when the impurity scattering is isotropic. \cite{berg84} Its
precise definition is \beq \frac{1}{\tau}=2\pi
n_{\rm{imp}}N(0)|v(\hat{p},\hat{p}')|^2, \label{timescat} \eeq
where $n_{\rm{imp}}$ is the impurity concentration, and
$v(\hat{p},\hat{p}')$ is the probability amplitude for an incoming
electron with momentum direction $\hat{p}'$ to be scattered into
direction $\hat{p}$ after collision with an impurity. Its square
modulus is assumed to be independent of both the incoming and the
outgoing direction. Finally, for a macroscopic description,
impurity scattering is characterized by the {\it disorder rate}
$\hbar/\tau$. We will therefore use
$\check{\Sigma}=\check{\Sigma}_{\rm{BCS}}+
\check{\Sigma}_{\rm{imp}}$ as the self-energy of our problem.

The inclusion of a finite superfluid velocity in our set of
equations leads to the addition of phase factors $e^{\pm i{\bf
q}\cdot {\bf R}}$ in some physical quantities. Their effect is
equivalent to that of shifting the energy variable by an amount
$\hbar v_F \hat{p}\cdot{\bf q}$ \cite{ramm86,ferr98b} in Eq.
(\ref{motioneq}) (${\bf q}$ is half the Cooper pair momentum,
$q=mv_s/\hbar$). The presence of ${\bf q}$ leads to a nonzero
supercurrent density and, through self-consistency equation
(\ref{deltaeq}), has a direct effect on the value of $|\Delta|$.
We wish then to solve the set of Eqs. (\ref{motioneq}),
(\ref{nc}), (\ref{curreq}), and (\ref{deltaeq}) for different
values of the superfluid velocity and $\hbar/\tau$.

If quasiparticles are in equilibrium with themselves and with
respect to the lattice, the Keldysh part of the Green's function
matrix (\ref{qcgf}) can be expressed in terms of the retarded and
advanced elements as
$\hat{g}=\tanh(E/2k_BT)(\hat{g}^R-\hat{g}^A)$. \cite{ramm86}
These, in turn, may be written as \beqa \hat{g}^R &=&
\alpha\hat{\tau}_3+\beta\hat{\tau}_1 \nonumber\\ \hat{g}^A &=&
-\alpha^*\hat{\tau}_3+\beta^*{\tau}_1, \label{gdos} \eeqa where
the scalar functions $\alpha \equiv \alpha({\bf R},\hat{p},E)$ and
$\beta \equiv \beta({\bf R},\hat{p},E)$ are the {\it generalized
densities of states}. \cite{ramm86} The normalization condition
(\ref{nc}) requires $\alpha^2+\beta^2=1$. If one considers a bulk
superconducting wire with periodic boundary conditions, the
problem becomes uniform and the relevant equation of motion
(\ref{motioneq}) yields \beq (E-\hbar v_F q u)\beta
-i|\Delta|\alpha +\frac{i\hbar}{2\tau}(\beta
\!<\!\alpha\!>\!-\alpha\!<\!\beta\!>\!)=0, \label{doseq} \eeq with
$u\equiv \hat{p}\cdot{\bf q}/q$ for the angular variable. This
equation is solved in Appendix \ref{averaged} with full inclusion
of the $u$-dependence. \cite{comm1} Once the functions $\alpha
(u,E)$ and $\beta (u,E)$ are obtained, we compute the value of
$|\Delta|$ from Eq. (\ref{deltaeq}). This new value is introduced
in (\ref{doseq}) to calculate again the generalized densities of
states, from which in turn we obtain a new order parameter. The
procedure is repeated until self-consistency in $|\Delta|$ is
achieved. After a self-consistent pair potential is found for
given superfluid velocity and disorder, the current density is
calculated with Eq. (\ref{curreq}).

\subsection{Discussion}

In Fig. \ref{dvsvs} the order parameter amplitude $|\Delta|$ is
plotted as a function of the superfluid velocity for different
temperatures, and dimensionless disorder strengths
$\Gamma\equiv\hbar/\tau\Delta_0$. One may note that Anderson's
theorem \cite{ande59} is satisfied at all temperatures, since
$|\Delta|$ for $v_s=0$ is independent of the disorder strength.
For very small $\Gamma$, one retrieves the expected result that
$|\Delta|\rightarrow 0$ when $v_s p_F/|\Delta| \sim 1$. Increasing
the disorder seems to reinforce the pair potential because of the
much larger values of $v_s$ needed to supress $\Delta$. This
occurs at $v_s=v_d$, with $v_dp_F/\Delta_0\simeq 3.1$, $2.6$, and
$1.3$ for $T/T_c=0.1$, $0.5$, and $0.9$, when $\Gamma=10$, i.e.,
more than twice the ballistic (small $\Gamma$) value. This
behavior may be interpreted as the tendency of disorder to restore
the spherical symmetry (and thus sustain the order parameter
amplitude), counteracting the anisotropy induced by the presence
of superflow (see following subsection).

This effect should not be viewed as an enhancement of
superconductivity by disorder. Inspection of Fig. \ref{jcvsvs}
shows that, as expected, the critical current density $j_c$
(defined as the maximum possible value of $j$ with a nonzero
$\Delta$) decays with increasing disorder. This is compatible with
the gap behavior discussed in the previous paragraph, because the
density of superfluid electrons is also reduced. \cite{dege66}
Fig. \ref{jcvsvs} also shows that, for a given macroscopic current
density $j$, there are two possible values of $v_s$. The smaller
$v_s$ yields the more stable configuration. Finally, we note that,
for $T$ close to $T_c$, we reproduce the smooth behavior expected
from Ginzburg-Landau calculations. \cite{dege66}

As an illustration, we plot in Fig. \ref{cmvsdt} the values of
several physical magnitudes at the point $j=j_c$ as a function of
disorder (left pannels) and temperature (right pannels). Despite
the above mentioned different depairing behavior, both critical
$j$ and $|\Delta|$ actually diminish with increasing disorder and
temperature. However, the critical superfluid velocity is enhanced
with $\Gamma$ while it decreases with $T$. This is a manifestation
of the high anisotropy needed to break superconductivity when
disorder is strong. It also shows that disorder is not
intrinsically depairing, while temperature is.

Finally, in Fig. \ref{jcvsd} we plot the decay of the critical
current density with disorder at low temperatures ($T=0.1T_c$) for
a wide range of $\Gamma$ values. Inset (a) shows the linear
behavior for small disorder ($\Gamma\!<\!0.1$). Inset (b)
indicates that for $\Gamma\!>\!10$ the critical current follows a
power-law ($j_c\propto \Gamma^{-\beta}$). Here we present a
unified treatment encompassing the ballistic and the diffusive
regimes. This requires going beyond the Usadel equations
\cite{usad70}, valid only within the dirty limit. Our results may
be compared with the predictions of a macroscopic theory based on
energetic arguments. If we use a phenomenological formula for the
density of superfluid electrons (valid for $l \ll \xi_0$, see Ref.
\cite{dege66}), \beq \rho_s(l)=\frac{\rho_s(\infty)}{1+\xi_0/l},
\label{supere} \eeq and equate the kinetic energy to the constant
condensation energy $j_c^2/2\rho_s$, the law
$j_c(\Gamma)=j_c(0)/(1+\Gamma)^{1/2}$ is obtained. This is the
dotted line in Fig. \ref{jcvsd}. This simple law cannot reproduce
the entire $\Gamma$ dependence, and, in particular, it yields an
exponent 0.5 slightly different from the exact $\beta\approx 0.47$
which we obtain numerically.

\subsection{Density of states}

In Fig. \ref{dos} we plot the density of states (DOS) for
different values of the superfluid velocity and the disorder
strength. At low values of $v_s$ ($\ll v_d$) we recover the
characteristic BCS density of states. The splitting of the gap
$\Delta$ into $\Delta_+$ and $\Delta_-$ modifies the quasiparticle
DOS. As $v_s$ raises from zero, the singularity at $E=\Delta_0$
evolves into two cusps at $\Delta_{\pm}$. The smoother character
of the split singularity comes from the fact that the minimum
quasiparticle energy depends on the momentum direction. Thus, the
zero-velocity DOS singularity actually evolves into a distribution
of singularities which, when integrated over the Fermi surface,
yields two characteristic cusps.

For weak and moderate disorder ($\Gamma=0.1$ and 1), the GS regime
can be achieved with values of $v_s$ within the stable branch (see
Fig. \ref{jcvsvs} for the corresponding temperature $T/T_c=0.1$).
This result is important, since it shows that the GS state occurs
in a stable manner for equilibrium transport in relatively clean
superconducting wires. GS can also be reached in dirty
($\Gamma=10$) superconductors but only for unstable values of
$v_s$.

Comparison of the DOS curves for $\Gamma=1$ and $\Gamma=10$ in
Fig. \ref{dos} shows that, for some values of $v_sp_F/\Delta_0$
(such as 1 or 1.5) for which one would expect to have GS, {\it the
effect of disorder is that of restoring the gap}. This effect may
be understood as resulting from the directional randomization
induced by multiple impurity scattering. The idea that GS should
exist for $v_s>|\Delta|/p_F$ comes from a kinematic analysis that
applies to plane waves or to quasiparticle states differing little
from them. This is the case of weak disorder. By contrast, in
strongly disordered superconductors, the exact quasiparticle
states necessarily involve a strong mixture of plane waves
pointing in many directions. Disorder helps to preserve the gap in
the exact DOS because, at very low energies, semiclassical
quasiparticle trajectories can only select plane waves from a
narrow solid angle of gapless directions. Not being able to mix
many momentum directions, quasiparticle states at such low
energies cease to exist. This effect is analogous to the
appearance of a minigap in the DOS of a diffusive normal metal in
contact with a superconductor. \cite{golu97,belz96}

\section{Nonequilibrium transport in a disordered wire}

The last section of this paper is devoted to nonequilibrium
transport in a disordered superconductor symmetrically connected
to two normal reservoirs with different chemical potentials, as
schematically depicted in Fig. \ref{nsnpic}. This boundary
condition introduces a nontrivial spatial dependence of the
physical magnitudes and makes the  equation of motion
(\ref{motioneq}) more difficult to solve. In particular, the
Keldysh component of the Green's function matrix (\ref{qcgf})
contains now all the relevant information on the quasiparticle
local distribution function and on the momentum-relaxation
processes taking place within the superconductor due to disorder.
Of course, the main new physical ingredient involved in our
calculation is the presence of a phase gradient in the order
parameter, which has to be determined self-consistently at each
point for every value of the chemical potentials at the
reservoirs. Like in Ref. \cite{sanc98}, the loss of translational
invariance makes it harder to compute the pair potential, which
now has to be determined both in its real and imaginary parts, as
demanded by a locally self-consistent calculation.

We are interested in the behavior of the new transport regimes
which were mentioned in the Introduction. In Refs.
\cite{sanc96,sanc97} it was shown how the regime dominated by AT
requires the presence of moderately reflecting barriers at the NS
contacts in order to be realized. Unfortunately, the mathematical
implementation of the appropriate boundary conditions in the
context of the QCGF is very cumbersome, since it involves the
solution of non-linear equations. \cite{comm2} Due to this
complication, we have chosen to perform the study for ideal NS
contacts (where the boundary conditions reduce to continuity of
the QCGF at the interfaces), leaving for the future a more
systematic calculation for arbitrary barriers at the interfaces,
in the spirit of Ref. \cite{sanc98}. With this assumption, the AT
regime cannot be studied, and we will focus instead on the
sensitivity of the GS regime to the degree of disorder in the
superconducting wire.

\subsection{Solution of the kinetic problem}

We assume a superconducting wire of length $L$, connected at its
ends $z=\pm L/2$ with perfect normal leads (characterized by
having $\Delta= 0$). These normal leads are connected to large
normal reservoirs through ideal contacts, so that the chemical
potentials characterizing the population of {\it incoming}
electrons and holes are those of the reservoirs from which they
were injected. \cite{sols99} Periodic boundary conditions are
considered in the transverse directions, perpendicular to the
transport direction $z$. For a given voltage $V=(\mu_L-\mu_R)/e$,
we wish to solve again Eq. (\ref{motioneq}) with the normalization
prescription (\ref{nc}) for a specific shape of the order
parameter $\Delta (z)\equiv |\Delta(z)|e^{i\varphi(z)}$. Once this
is done, the solution is used to compute a new $\Delta(z)$, and
the procedure is repeated until self-consistency is achieved.
Finally, the current density $j$ is calculated from Eq.
(\ref{curreq}).

In the stationary limit, and for an arbitrary shape of $\Delta$,
it is possible to parametrize the retarded and advanced QCGF as
\cite{ramm86} \beqa \hat{g}^{R} &=&
\alpha\hat{\tau}_3+\beta\hat{\tau}_1+\gamma\hat{\tau}_2
\nonumber\\ \hat{g}^{A} &=&
-\alpha^*\hat{\tau}_3+\beta^*\hat{\tau}_1+\gamma^*\hat{\tau}_2,
\label{zgra} \eeqa where $\alpha$, $\beta$, and $\gamma$, are now
scalar functions of the longitudinal coordinate $z$ (as well as of
$u$ and $E$). When Eq. (\ref{zgra}) is introduced into the
equation of motion (\ref{motioneq}) a new set of equations is
obtained: \beqa \hbar uv_F\frac{\partial\alpha}{\partial
z}&=&(i\frac{\hbar}{\tau}\!<\!\gamma\!>\!-2i{\mbox{Im}}\Delta)
\beta-(i\frac{\hbar}{\tau}\!<\!\beta\!>\!+2i{\mbox{Re}}\Delta)\gamma
\nonumber\\ \hbar uv_F\frac{\partial\beta}{\partial
z}&=&(-i\frac{\hbar}{\tau}\!<\!\gamma\!>\!+2i{\mbox{Im}}\Delta)
\alpha+(i\frac{\hbar}{\tau}\!<\!\alpha\!>\!+2E)\gamma \nonumber\\
\hbar uv_F\frac{\partial\gamma}{\partial
z}&=&(i\frac{\hbar}{\tau}\!<\!\beta\!>\!+2i{\mbox{Re}}\Delta)
\alpha-(i\frac{\hbar}{\tau}\!<\!\alpha\!>\!+2E)\beta, \label{zabc}
\eeqa where, as usual, the brackets stand for angular average. The
normalization condition (\ref{nc}) implies
$\alpha^2+\beta^2+\gamma^2=1$. In Appendix \ref{s-m} we solve this
set of equations with the appropriate boundary conditions.

While the retarded and advanced parts of the equation of motion
still give us the generalized densities of states of the problem,
the Keldysh part $\hat{g}$ of (\ref{qcgf}), which contains
information on actual occupations, is no longer trivial. Being in
a nonequilibrium context, we need to write its corresponding
equation of motion, which reads \beq \hbar
uv_F\frac{\partial\hat{g}}{\partial z}=iE(\hat{\tau}_3\hat{g}-
\hat{g}\hat{\tau}_3)-i(\hat{\Sigma}^R\hat{g}+\hat{\Sigma}\hat{g}^A-
\hat{g}^R\hat{\Sigma}-\hat{g}\hat{\Sigma}^A). \label{zg} \eeq One
way of parametrising $\hat{g}$ which automatically satisfies Eq.
(\ref{nc}) is \beq \hat{g}=\hat{g}^R\hat{h}-\hat{h}\hat{g}^A,
\label{schmscoe} \eeq with $\hat{h}$ an arbitrary distribution
matrix. Schmid and Sch\"on \cite{schm75} proposed a distribution
matrix $\hat{h}$ of the diagonal form
$\hat{h}=(1-2f_L)-2f_T\hat{\tau}_3$. This procedure is extremely
useful in situations where the dirty limit is valid and the
boundary conditions are independent of $u$, \cite{lamb97,golu97}
but it does not permit a complete separation of the densities of
states and the quasiparticle distribution function. \cite{beye85}
An alternative parametrization was suggested by Shelankov,
\cite{shel80} \beq \hat{g}=\frac{1}{2}\left
(f_1\hat{P}_1^R\hat{P}_1^A+f_2\hat{P}_2^R\hat{P}_2 \right),
\label{zparg} \eeq where $f_{1,2}$ are scalar functions of the
variables $(z,u,E)$, and $\hat{P}_{1,2}^R\equiv 1\pm\hat{g}^R$ and
$\hat{P}_{1,2}^A\equiv 1\mp\hat{g}^A$ satisfy \beqa
\hat{P}_i^{R(A)}\hat{P}_i^{R(A)}&=&2\hat{P}_i^{R(A)} \;\;\; i=1,2
\nonumber \\ \hat{P}_1^{R(A)}\hat{P}_2^{R(A)}&=&0.
\label{proyectors} \eeqa Multiplying Eq. (\ref{zg}) by
$\hat{P}_1^{A}\hat{P}_1^{R}$ on the left, taking the trace, and
using the cyclic properties as well as Eq. (\ref{proyectors}), one
may find after some algebra, \beq \hbar
uv_F{\mbox{Tr}}(\hat{P}_1^{A}\hat{P}_1^{R})\frac{\partial
f_1}{\partial z}=
-i{\mbox{Tr}}\{\hat{P}_1^{A}\hat{P}_1^{R}[(\hat{\Sigma}^R-\hat{\Sigma}^A)
f_1-\hat{\Sigma}] \}. \label{zf} \eeq Eq. (\ref{zf}) is a closed
expression that permits to obtain $f_1(z,u,E)$ since, due to the
symmetry of the problem, $f_2(z,u,E)= -f_1(-z,u,E)$. Actually,
$(1-f_1)/2$ coincides with the usual Fermi distribution function
in equilibrium, and with the quasiparticle distribution function
within a semiconductor model. \cite{beye85} To fix the population
of incoming electrons from each reservoir, it is natural to use
the boundary conditions \beqa f_1(-L/2,u,E)=\tanh[(E-eV/2)/2k_BT],
\;\;\; {\mbox{if}} \; u\!>\!0 \nonumber \\
f_1(L/2,u,E)=\tanh[(E+eV/2)/2k_BT], \;\;\; {\mbox{if}} \; u\!<\!0.
\label{zfbc} \eeqa

The resolution of Eq. (\ref{zf}) with this boundary conditions
also requires a {\it self-consistent determination of
$\check{\Sigma}$}. The contribution of $\check{\Sigma}_{\rm{imp}}$
has to be calculated at each energy from the solutions of Eqs.
(\ref{zabc}) and (\ref{zf}). The corresponding angular average of
the whole $\check{g}$ is then performed and introduced in Eq.
(\ref{sigmaimp}) until one achieves a self-consistent value of
$\check{\Sigma}_{\rm{imp}}$. Once this is done at every energy
$E$, the self-consistent procedure for $\Delta$ may be initiated.

Finally, when the self-consistent solutions for a given voltage
are found, one may calculate the electrostatic potential in the
structure as \cite{ramm86} \beq e\phi (z)=-\frac{1}{4}\int
dE\!<\!{\mbox{Tr}}\hat{g}\!>\!, \label{zphi} \eeq which is
directly related to the electronic density. Eqs. (\ref{zphi}) and
(\ref{curreq}) will be used to compute the electrostatic potential
profile and the current density which we discuss in the next
subsection.

\subsection{Discussion}

In Fig. \ref{jdvsv} we plot the variation of the current density
with the applied bias $V$ between the normal reservoirs for
different values of the disorder. The variation of the order
parameter amplitude at the center of the NSN structure $z=0$ is
represented in the lower figure. Throughout this subsection, we
use the values of $L=3\xi_0$ for the length of the superconductor
and $T=0.01T_c$ for the temperature. One may see how when disorder
is present ($\Gamma \neq 0$), {\it there exist two regimes with
different current-voltage slopes}. For low bias, $dj/dV$ is always
bigger than its corresponding normal value (which is that attained
at higher voltages). This fact shows clearly the different effect
of diffusive (dirty) regions and tunnel junctions on
superconducting transport. In a purely ballistic conductor
attached via tunnel barriers to the reservoirs, the conductance
would always be bigger in the normal state, \cite{hui93} the main
reason being the much smaller probability for the simultaneous
tunneling of two electrons forming a Cooper pair. However, if the
contacts are good enough, and scattering is dominated by spatially
distributed impurities, the superconducting state supports a
greater amount of current due to the lack of normal reflection of
quasiparticles decaying into Cooper pairs. In a perfectly clean
sample, the superconducting and the normal states cannot be
distinguished in what regards to transport (both display the same
slope in the $\Gamma=0$ case). \cite{sanc95,mart95,sanc97,sols99}

There is a correlation between the slope discontinuity for the
current and the vanishing of $\Delta(z=0)$. This is particularly
clear for $\Gamma \agt 1$ since then the effect is more marked and
occurs at lower voltages ($eV \simeq 1.3 \Delta_0$). This is the
point at which the whole structure becomes normal and transport is
governed by the Boltzmann equation. The case of weak disorder
($\Gamma=0.2$) is a bit different. One may notice that the current
density goes down slightly above $eV=1.5\Delta_0$, while
$\Delta(0)$ remains finite for still higher voltages. {\it The
structure is still superconducting} (in the sense of having a
nonzero order parameter) but the current is practically
indistinguishable from that of the normal state. {\it The
superconductor is in the GS regime}
\cite{sanc95,mart95,sanc97,sanc98} (see discussion below).

In Fig. \ref{vddnvsz} one may compare in more detail the
differences between the situations with weak and strong disorder.
We will concentrate on the different behavior of the GS regime.
Several physical magnitudes have been represented as a function of
position for different values of the applied voltage. The top
pannels represent the electrostatic potential along the
superconductor S. Due to the interfaces with normal reservoirs and
to the presence of an applied voltage, there is a penetration of
the electric field within S, causing the quasiparticles to have a
chemical potential different from that of the condensate (equal to
zero in these graphs). \cite{clar72}

The relaxation of the electric field within a superconductor is
one of the most heavily studied topics in the literature on
nonequilibrium superconductivity. \cite{gray81} The physics we
encounter here is somewhat different because we deal with a finite
superconductor without inelastic scattering, connected
symmetrically to normal reservoirs. Two different elastic
mechanisms contribute to the spatial relaxation, after length
$\Lambda$, of the excess charge density due to quasiparticles near
the boundaries. One may have normal reflection of quasiparticles
returning to the reservoir. This translates into a law
$[1+(z+L/2)/l]^{-1}$ for $\phi(z)$ on the left boundary,
characteristic of decay by incoherent multiple normal scattering.
\cite{sols99,datt95} On the other hand, beyond a certain length,
quasiparticles cannot propagate because their energy is
insufficient to overcome the effective gap, since this increases
as one gets deeper into the superconductor. This occurs at a
characteristic length from the boundary $\hbar v_F/\epsilon(V)$,
where $\epsilon(V)$ is an effective energy that diminishes with
$V$ because of the self-consistent decrease of $|\Delta|$ in all
the structure. This is the length scale at which quasiparticle
conversion into Cooper pairs by Andreev reflection takes place.
Andreev scattering generates a quasiparticle of opposite charge
and thus tends to quickly suppress the charge excess.
\cite{sols99} These considerations give a good account of the
results for $\phi(z)$ in Fig. \ref{vddnvsz}, which shows a roughly
constant $\Lambda$ as a function of the applied voltage up to the
transition to the normal state for $\Gamma=5$ (dirty limit, $l$
dominates), and a continuously increasing $\Lambda$ when
$\Gamma=0.2$ (clean limit, $\hbar v_F/\epsilon(V)$ dominates).
When the structure becomes normal, the voltage profile is that
which results from substracting the curves $[1+(L/2\pm
z)/l]^{-1}$. In the very clean limit ($l \gg L$) this results in
an essentially straight line.

Center and bottom pannels represent the $|\Delta|$ and $\Delta_-$
profiles, respectively. Since $\varphi '(z)$ is no longer
constant, $\Delta_-(z)\equiv |\Delta(z)|-\hbar v_F \varphi '(z)
/2$. When $\Delta_-\!<\!0$ at some point, the superconductor
becomes locally gapless, at least for plane waves. \cite{comm4} In
particular, near the interfaces, $\Delta_-$ becomes negative very
quickly as $V$ increases, due to the smallness of $|\Delta|$,
while it remains positive in the bulk (near $z=0$). As the voltage
increases, one may note a different evolution of the left and
right graphs before the normal state is reached (this occurs at
$eV=1.86\Delta_0$ and $eV=1.35\Delta_0$, respectively). {\it In
the diffusive regime, the superconductor cannot be globally
gapless}. All the $\Delta_-$ curves are positive in the central
region. Quasiparticles with energies below the effective gap
penetrate the superconductor partially and become Cooper pairs.
However, if $\Delta_-(0)\!<\!0$, this transfer process becomes
marginal. Normal scattering dominates and, due to the strong
disorder, the quasiparticles can hardly be transmitted across the
system. The directional randomization \cite{sanc97} results in a
sharp transition to the normal state. On the contrary, for
$\Gamma=0.2$, Fig. \ref{vddnvsz} shows that it is still possible
to have $\Delta_-\!<\!0$ for all $z$ (for, e.g.,
$eV=1.5\Delta_0$), while the system is still superconducting. The
quasiparticle density of states is nonzero at all energies because
the weak disorder is unable to randomize the quasiparticle
nonequilibrium distribution, which results in the preservation of
superconductivity. \cite{sanc97} For sufficiently high voltages
(depending on the wire length and on its effective dimensionality
\cite{roge60}), the GS becomes unstable and the system goes
normal.

Since realistically clean superconducting wires can be obtained
nowadays, it could be possible to measure finite superflow effects
passing a large current along a given direction of a
superconducting sample, while probing the gap and/or the density
of states with a weak tunnelling current.

\section{Conclusions}

We have used the technique of the quasiclassical Green's function
(QCGF) to study the effect of disorder and many transport channels
in situations where superconductors present a non-negligible
superfluid velocity. The QCGF technique has allowed us to include
impurity scattering in a tractable manner, and to perform
realistic calculations of disorder-averaged physical quantities.

The first part of this article has been devoted to the study of
equilibrium transport in homogeneous wires. We have seen that
disorder tends to restore the spherical symmetry and strengthen
the order parameter amplitude, in contrast with the pair breaking
effect of finite condensate flow. However, its net effect is that
of diminishing the current. The critical current density has been
calculated for disorder strengths ranging from the clean to the
dirty limit. The self-consistent density of states for different
values of the superfluid velocity and disorder has also been
discussed. We find that, due to its direction mixing effect,
sufficient disorder restores the quasiparticle gap in transport
contexts where one would expect gapless superconductivity (GS).

The second part has focussed on nonequilibrium transport in
normal-superconductor-normal structures. Clean contacts between
the superconductor and the normal reservoirs have been assumed,
with boundary conditions that describe incoming quasiparticles
from reservoirs with different chemical potentials. We have found
that the current-induced GS regime is very sensitive to the
presence of disorder. It is suppressed in the diffusive limit, but
is stable in sufficiently clean superconductors. We have
calculated the spatial profiles of the electrostatic potential,
order parameter amplitude, and energy threshold for quasiparticle
transmission. We have found that they show subtle differences in
the ballistic and diffusive regimes in what regards to the
current-induced phase transition to the normal state.

\acknowledgments

We wish to thank W. Belzig, J. Ferrer, C.J. Lambert, A.J. Leggett,
R. Raimondi, G. Sch\"on, A.F. Volkov, and A.D. Zaikin, for
valuable discussions. This work has been supported by the
Direcci\'on General de Investigaci\'on Cient\'{\i}fica y T\'ecnica
under Grant No. PB96-0080-C02, and by the TMR Programme of the EU.

\appendix

\section{Impurity-averaged generalized densities of states} \label{averaged}

Eq. (\ref{doseq}) is actually an algebraic equation if one
considers the averaged $c(E)\equiv <\!\alpha(u,E\!>\!$ and
$s(E)\equiv <\!\beta(u,E)\!>\!$ as known quantities. If it is so,
one may define $t(u,E)\equiv \beta/\alpha$, whose formal solution
is \beq t(u,E)=\frac{i(|\Delta| + \hbar s(E)/2\tau)}{(E-\hbar v_F
q u+ i\hbar c(E)/2\tau)}. \label{tsol} \eeq

Now, $c(E)$ and $s(E)$ may be found if one notices the
normalization prescription $\alpha^2+\beta^2=1$, which makes
$\alpha=(1+t^2)^{-1/2}$ and $\beta=t(1+t^2)^{-1/2}$. If we change
variables in the angular integrals for $c(E)$ and $s(E)$: \beqa
c(E) &=& \int_{t_+}^{t^-}\frac{i(|\Delta|+ \hbar
s(E)/2\tau)}{2\hbar v_F q}\frac{dt}{t^2(1+t^2)^{1/2}} \nonumber
\\ s(E) &=& \int_{t_+}^{t^-}\frac{i(|\Delta|+
\hbar s(E)/2\tau)}{2\hbar v_F q}\frac{dt}{t(1+t^2)^{1/2}},
\label{csint} \eeqa with $t_{\pm}(E)\equiv i(|\Delta|+\hbar
s(E)/2\tau)/(E\pm\hbar v_F q +i\hbar c(E)/2\tau)$. Integrals in
Eq. (\ref{csint}) give two coupled equations for $c(E)$ and $s(E)$
\beqa c&=&\frac{i(|\Delta| + \hbar s/2\tau)}{2\hbar v_F
q}\left[\frac{(1+t_+^2)^{1/2}}{t_+}
-\frac{(1+t_-^2)^{1/2}}{t_-}\right] \nonumber\\
s&=&\frac{i(|\Delta| + \hbar s/2\tau)}{2\hbar v_F
q}\log\frac{t_-[1+(1+t_+^2)^{1/2}]} {t_+[1+(1+t_-^2)^{1/2}]},
\label{csselfeq} \eeqa where we have omitted the dependence on $E$
for simplicity. These equations can be solved self-consistently at
each energy using standard techniques.

\section{The Schopohl-Maki transformation} \label{s-m}

Eqs. (\ref{zabc}) can be decoupled using the Schopohl-Maki
transformation \cite{scho98,belz98} \beq y_{1,2}=\frac{\beta\mp
i\gamma}{1+\alpha}, \label{ydef} \eeq which leads, with the help
of Eq. (\ref{nc}), to the following Riccati differential
equations: \beqa \hbar uv_F\frac{\partial y_1}{\partial
z}-2i\tilde{E}y_1+\tilde{\Delta}_2 y_1^2-\tilde{\Delta}_1&=&0
\nonumber\\ \hbar uv_F\frac{\partial y_2}{\partial
z}+2i\tilde{E}y_2-\tilde{\Delta}_1 y_1^2+\tilde{\Delta}_2&=&0,
\label{riccati} \eeqa where $\tilde{E}\equiv
E+i\hbar\!<\!\alpha\!>\!/2\tau$, $\tilde{\Delta}_1\equiv
\Delta+\hbar\!<\!\beta-i\gamma\!>\!/2\tau$, and
$\tilde{\Delta}_2\equiv
\Delta^*+\hbar\!<\!\beta+i\gamma\!>\!/2\tau$.

When one solves Eq. (\ref{riccati}) in the normal leads (where
$\Delta=0$ and $\Gamma=0$), the solutions are \beqa
y_1&=&y_1(z_0)e^{2iE(z-z_0)/\hbar uv_F} \nonumber\\
y_2&=&y_2(z_0)e^{-2iE(z-z_0)/\hbar uv_F}, \label{normalsols} \eeqa
with $z_0\equiv\pm L/2$. In the imaginary time representation, one
has the freedom of choosing either the positive or the negative
imaginary axis for the Matsubara energies involved in the problem.
Considering the usual criterion of $E=i\omega_n$, with
$\omega_n\!>\!0$, one obtains as appropriate boundary conditions
that ensure the finiteness of the solutions in the normal leads
\cite{brud90} \beqa y_1(-L/2)=y_2(L/2)=0 \;\;\; {\mbox{if}} \;
u\!>\!0 \nonumber\\ y_2(-L/2)=y_1(L/2)=0 \;\;\; {\mbox{if}} \;
u\!<\!0. \label{ybc} \eeqa On the other hand, the symmetry
relation $\Delta(z)=\Delta^*(-z)$ guarantees the following
symmetries for the solutions of the generalized densities of
states (regardless of the voltage difference between the normal
reservoirs): \beqa \alpha(z,u,E)&=&\alpha(-z,u,E)\nonumber\\
\beta(z,u,E)&=&\beta(-z,u,E)\nonumber\\
\gamma(z,u,E)&=&-\gamma(-z,u,E), \label{abcsym} \eeqa which
translates into \beq y_1(z,u,E)=y_2(-z,u,E). \label{ysym} \eeq Eq.
(\ref{riccati}) reduces then to one effective equation for each
$y$, which can be solved numerically, and the generalized
densities of states may finally be calculated from the relations
\beqa
\alpha(z)&=&\frac{1-y_1(z)y_1(-z)}{1+y_1(z)y_1(-z)}\nonumber\\
\beta(z)&=&\frac{y_1(z)+y_1(-z)}{1+y_1(z)y_1(-z)}\nonumber\\
\gamma(z)&=&\frac{i[y_1(z)-y_1(-z)]}{1+y_1(z)y_1(-z)},
\label{finalabc} \eeqa where we have omitted the $u$ and $E$
dependence. As explained at the end of subsection IV.A, a
self-consistent procedure is also needed at each energy to obtain
the right angular averages entering the equations of motion.

\dvsvs \jcvsvs \cmvsdt \jcvsd \dos \nsnpic \jdvsv \vddnvsz


\begin{references}
\bibitem{comm1} In presence of superflow, the anisotropy associated
with the moving condensate becomes important, and the customary
use of the Usadel equations \cite{usad70} is no longer safe.
Specifically, Usadel equations are useful in contexts where the
current is small and may be calculated via the second term in the
angular expansion $\check{g}(u)\simeq\check{g}_0+u\check{g}_1+...$
\cite{lamb97} However, the general expression for the current
given in Eq. (\ref{curreq}) needs, in principle, the full
$u$-dependence of the Green's function.
\bibitem{comm2} Boundary conditions have been studied in a systematic way by
Zaitsev \cite{zait84}, who derived a generic expression for the
case of an arbitrary interface. In Ref. \cite{lamb97} one may find
a derivation of the procedure to obtain the right boundary
conditions starting from the scattering formalism.
\bibitem{comm4} Here, the kinematic criterion $v_s >
|\Delta|/p_F$ coincides in practice with the exact DOS prediction.
For low disorder, both criteria coincide (see subsection III.C),
while for strong disorder we will see that GS does not exist.
\end{references}
\end{document}